# Cross-polarized optical absorption of single-walled nanotubes probed by polarized photoluminescence excitation spectroscopy

Yuhei Miyauchi, Mototeru Oba and Shigeo Maruyama*

Department of Mechanical Engineering, The University of Tokyo, 7-3-1 Hongo, Bunkyo-ku, Tokyo 113-8656, Japan

**Abstract**

Cross-polarized absorption peaks of isolated single-walled carbon nanotubes were observed by a polarized photoluminescence excitation (PLE) spectroscopy. Using a simple theory for PL anisotropy, the observed PLE spectra are decomposed into 'pure' components of the photoexcitation for incident light polarized parallel and perpendicular to the SWNT axis. For several (n, m) SWNTs, distinct peaks corresponding to perpendicular excitation were observed. The measured transition energies for perpendicular excitations were blue-shifted compared to the qualitative values predicted within a single-particle theory. The results indicate a smaller exciton binding energy for perpendicular excitations than for parallel excitations.

*Corresponding author.
Electronic address: maruyama@photon.t.u-tokyo.ac.jp

## I. INTRODUCTION

Recently there has been considerable interest in the optical properties of single-walled carbon nanotubes (SWNTs) as one of the ideal quasi one-dimensional systems.[1] Due to their one-dimensionality, the optical responses of SWNTs are strongly dependent on the polarization direction of the incident light with respect to the SWNT axis.[2-4] Over the past few years, photoluminescence excitation (PLE) spectroscopy has been extensively used to probe the electronic structures of semiconducting SWNTs.[5-18] Theoretical studies and recent experiments have demonstrated that these optical transitions are dominated by strongly correlated electron-hole states in the form of excitons.[12-14,19-21] These PLE studies have mainly focused on the optical transitions for incident light parallel to the SWNT axis, and there have been a few for cross-polarized transitions. So far, Lebedkin et al.[7] have reported excitation and emission energy dependence of PL anisotropy for micelle-suspended SWNTs[5] synthesized by the laser vaporization method. They have measured PL anisotropy spectra at several excitation energies. Although no resolvable feature for the perpendicular excitation was observed, they found that contributions of the perpendicular transition moment to the photoexcitation depend on the excitation energies. In our previous study,[22] we have observed polarization-dependent relative enhancement of a PL peak for (7, 5) SWNTs in the polarized-PLE experiment on partially aligned SWNTs in a gelatin film, after clearly distinguished the phonon sideband PLE peaks by SW$^{13}$CNTs experiment.[18] For optical absorption[23] and Raman spectroscopy[24,25] using bundled SWNTs, there have been some experimental studies indicating the existence of the perpendicular excitation. However, the optical responses of SWNT bundles are completely different from those for SWNTs dispersed in a surfactant suspension[5] due to intertube interactions. Since such micelle-suspended SWNTs are frequently used for PL studies,[5-10, 17, 18] it is very important to investigate the anisotropic optical properties of isolated SWNTs in aqueous solutions for various (n ,m) types.

In this Letter, we present that distinct absorption peaks can be observed in the PLE spectra for the polarization perpendicular to the SWNT axis, although the perpendicular excitation has been considered to be strongly suppressed due to the induced self-consistent local field (depolarization effect[2]). Furthermore, using a procedure to determine the fractional contribution of parallel and perpendicular absorption and emission dipoles[26], the PLE spectra are decomposed into 'pure' components for parallel and perpendicularly polarized excitations. The observed transition energies for cross-polarized condition was considerably higher than theoretical predictions within a single particle theory. This discrepancy is discussed with the excitonic effect.

## II. EXPERIMENTAL
### A. Preparation of dispersed SWNTs in a surfactant suspension

In order to measure PL spectra from individual SWNTs in a surfactant suspension[5], SWNTs synthesized using the alcohol CCVD method[27] (supplied by Toray Industries, Inc.) were dispersed in D$_2$O with 0.5 wt % sodium dodecylbenzene sulfonate (NaDDBS)[8] by heavy sonication with an ultrasonic processor (Hielscher GmbH, UP-400S with H3/Micro Tip 3) for 1 h at a power flux level of 460 W/cm$^2$. These suspensions were then centrifuged (Hitachi Koki himac CS120GX with a S100AT6 angle rotor) for 1 h at 386 000 $g$ and the supernatants, rich in isolated SWNTs, were used in the PL measurements.

**B. Selection rules of optical transitions in SWNTs**

Figure 1 (a) and 1(b) schematically shows the selection rules for incident light polarized parallel and perpendicular to the SWNT axis.[2-4] In the case of parallel polarization, optical absorption between subbands with the same quasi-angular momentum are allowed ($\Delta\mu = 0$ or $E_{ii}$ transitions), while $\Delta\mu = \pm 1$ transitions are allowed for perpendicular polarization ($\mu$ denotes the subband (cutting line) index in the 2D Brillouin zone of graphite[1]). In this letter, we refer to $\Delta\mu = \pm 1$ transitions between the first and second subbands as $E_{12}$ and $E_{21}$ transitions. Since the emission is primarily caused by recombination of electrons and holes within the first conduction and valence subbands with the same index $\mu$ ($\Delta\mu = 0$, $E_{11}$ transitions), the emission dipole is considered to be parallel to the SWNT axis.

**C. Polarized PLE spectroscopy**

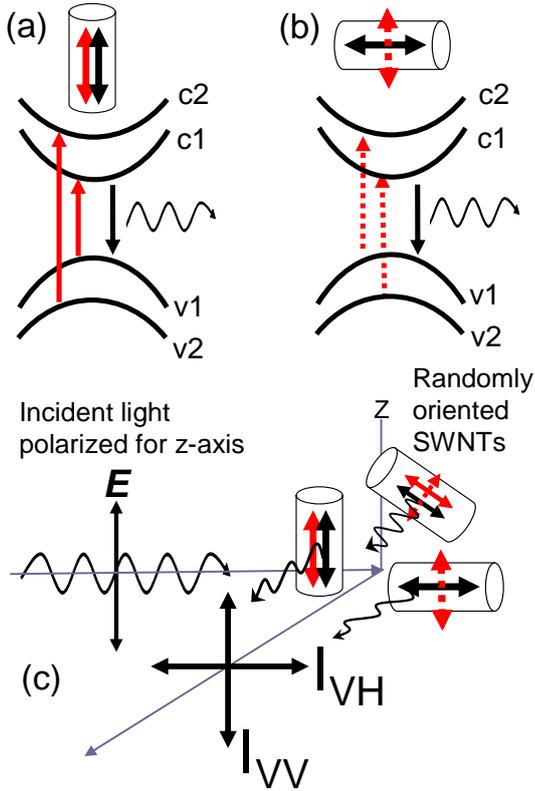

FIG. 1. (Color online) Schematics of optical transitions in SWNTs corresponding to (a) collinear and (b) perpendicular absorption and emission dipoles. Solid and dotted arrows indicate $\Delta\mu = 0$, and $\Delta\mu = \pm 1$ transitions, respectively. (c) Schematic diagram for the measurement of PL anisotropies of randomly oriented SWNTs.

Figure 1(c) schematically shows the configuration of our polarized PLE experiment. The sample is excited with vertically polarized light, and the emission is observed through another polarizer. The emission polarizer is oriented parallel to ($I_{VV}$) or perpendicular to ($I_{VH}$) the direction of the polarized excitation (z-axis). The measured spectral data were corrected for wavelength-dependent variations in excitation intensity and polarization dependent detection sensitivity.

Because a PL emission signal is enhanced when directions of an emission dipole and the emission polarizer are the same, the direction of emission polarizer determines the mainly probed SWNTs. In the $I_{VV}$ spectra, the PL emission excited by collinear absorption ($\Delta\mu = 0$ absorption → $\Delta\mu = 0$ emission) is enhanced, whereas emission excited by perpendicular absorption ($\Delta\mu = \pm 1$ absorption → $\Delta\mu = 0$ emission) is enhanced in the $I_{VH}$ spectra.

Near infrared emission from the sample was recorded while the excitation wavelength was scanned from the UV to NIR range. The excitation and emission spectral slit widths were both 10 nm and scan steps were 5 nm on both axes. The PL spectra were measured with a HORIBA SPEX Fluorolog-3-11 spectrofluorometer with a liquid-nitrogen-cooled InGaAs near-IR detector. A UV-VIS-NIR polarizer and a NIR polarizer were set behind the excitation monochromator and before the emission monochromator, respectively. The alignment of the polarizers was examined by observing the polarization of the scattered light from dilute colloidal silica in water.[26]

**III. RESULTS AND DISCUSSIONS**

**A. Observation of a distinct cross-polarized absorption peak**

Figure 2(a) and 2(b) shows the measured PL contour maps for $I_{VV}$ and $I_{VH}$ configurations, respectively. For the $I_{VH}$ configuration, the maximum PL intensities of the $E_{22}$ peaks are less than half of those in the $I_{VV}$ spectra. This result is due to the excitation photoselection[26] by the incident light polarized along the vertical (z) axis. Fig.2(d) compares the PLE spectra of (7, 5) for $I_{VV}$ and $I_{VH}$ configurations. As shown in Fig.2(b) and 2(d), the relative intensity of the peak marked by an asterisk (*) is enhanced in the $I_{VH}$ spectra. One can find similar features of (7, 5) SWNTs in previous reports of PL mapping by different groups using different SWNT samples.[6,9,10,18] In our previous isotopic PLE study using carbon-13 SWNTs,[18] we observed no distinct isotopic shift for this small peak of (7, 5) SWNTs. In addition, the result for the $I_{VH}$ configuration is consistent with our previous polarized PLE observations on partially aligned SWNTs in a gelatin film,[22] confirming the marked peak (*) in Fig. 2 is due to cross-polarized absorption.

**B. PL anisotropy**

For more quantitative treatments, the PL anisotropy $r(E_{ex}, E_{em})$ can be calculated from the measured $I_{VV}$ and

$I_{VH}$ spectra as[26]

$$r \equiv \frac{I_{VV} - I_{VH}}{I_{VV} + 2I_{VH}}, \quad (1)$$

where the denominator is defined as the total intensity $I_T \equiv I_{VV} + 2I_{VH}$. Assuming that SWNTs have only collinear (perpendicular) absorption and emission dipole moments, the maximum (minimum) anisotropy $r_{//}$ ($r_\perp$) is given as $r_{//} = 0.4$ ($r_\perp = -0.2$),[26] which is the maximum (minimum) value of $r$ for randomly oriented molecules.[26] The anisotropy for perpendicular transition moments $r_\perp$ can be calculated from $r_{//}$ as[26]

$$r_\perp = -0.5 r_{//}. \quad (2)$$

Since the anisotropy $r$ is expected to be large (small) when the absorption and emission dipoles are parallel (perpendicular), the value of the anisotropy can be used to estimate the contribution of respective dipoles for excitation at a given energy. Fig.2(c) shows contour maps of the calculated anisotropy $r_{exp}$ in the measured energy range, which are directly obtained using Eq.(1) from measured $I_{VV}$ and $I_{VH}$ shown in Fig.2 (a) and 2(b). It can be clearly seen that the observed anisotropy $r_{exp}$ obtains peak values at the $E_{22}$ excitation energies. The values of $r_{exp}$ at the $E_{22}$ energies were almost identical (~0.3) for different (n, m) SWNTs.

One can also find small $r_{exp}$ regions in the PLE spectra for respective (n, m) species. One of these small $r_{exp}$ areas, indicated by an oval mark in Fig. 2(c) corresponds to the peak marked with an asterisk (*) in Fig. 2(b) and 2(d). This result suggests that the contribution of parallel and perpendicular dipoles varies depending on the excitation energies.

## C. Decomposition of PL maps

At each excitation and emission energy, the measured anisotropy $r_{exp}$ can be decomposed using the fundamental anisotropies for parallel ($r_{//}$) and perpendicular ($r_\perp$) absorption and emission dipoles as $r_{exp} = f_{//} r_{//} + f_\perp r_\perp$, where $f_{//}$ and $f_\perp$ is the fractional contribution of parallel and perpendicular dipoles to the total intensity $I_T$, respectively.[26] Since the total fractional contribution is unity $f_{//} + f_\perp = 1$, the "pure" PL maps for parallel ($I_{//}$) and perpendicular ($I_\perp$) dipoles can be calculated using the relationships[26]

$$I_{//} = f_{//} I_T = \left( \frac{r_{exp} - r_\perp}{r_{//} - r_\perp} \right) I_T, \quad (3)$$

$$I_\perp = f_\perp I_T = \left( \frac{r_{//} - r_{exp}}{r_{//} - r_\perp} \right) I_T \quad (4)$$

Here, it is assumed that $r_{//}$ and $r_\perp$ are independent of the excitation energy and nanotube species, and that emission occurs only from the $E_{11}$ state.

In the real case, some possible processes can be expected to reduce the anisotropy by light scattering, reabsorption, radiationless energy transfer, rotational diffusion, and bending of nanotubes, although it is difficult to identify the precise cause(s) in this study. In order to take these possible reductions of anisotropy into account, we assigned the maximum value of observed anisotropy for (7, 5) SWNTs in the lower excitation energy range to the value of $r_{//}$ for the decomposition, since collinear excitation and emission is expected to be dominant near the $E_{11}$ transition energy. We selected $r_{//} = 0.31$ within the experimental error $r_{max} = 0.30 \pm 0.02$ in the analysis, since this value gave the best separation of $I_{//}$ and $I_\perp$ peaks throughout the measuring range. $r_\perp$ can be calculated by substituting $r_{//} = 0.31$ into Eq.(2). The maximum measured anisotropy is also in good agreement with the result of a previous anisotropy study (reporting $r_{max} = 0.32$ in the lower excitation energy region) of micelle-suspended SWNTs with thicker diameters[7]. Similar assignment of the maximum observed anisotropy to $r_{//}$ has been used for the decomposition of absorption spectra of indole[26].

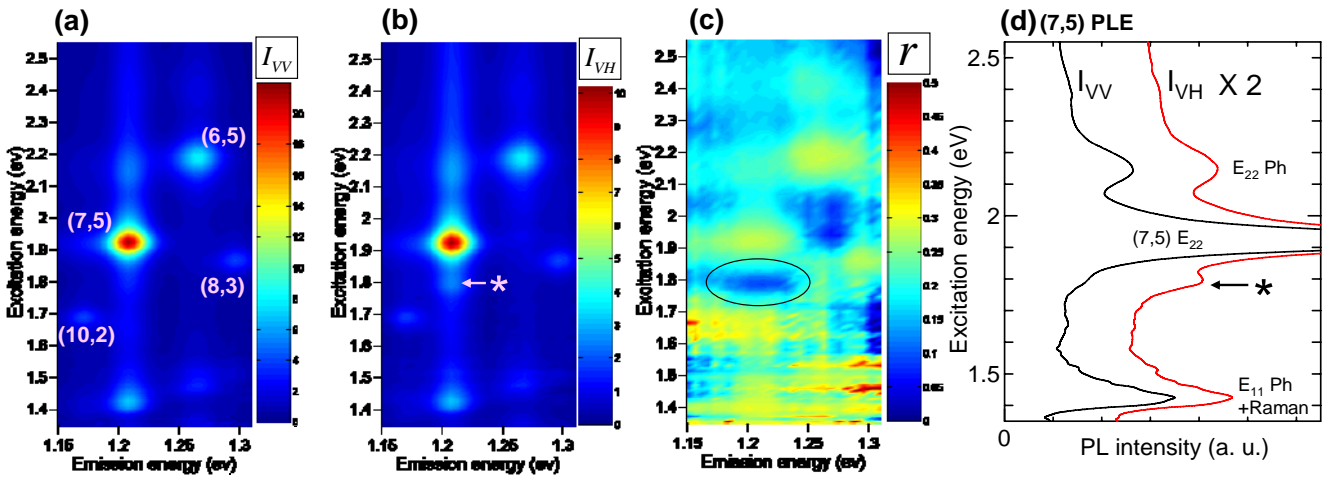

FIG. 2. (Color online) (a, b) PL and (c) anisotropy maps in the emission energy range of 1.15-1.31 eV. (d) PLE spectra of (7, 5) SWNTs for $I_{VV}$ and $I_{VH}$ configurations. The $I_{VH}$ spectrum in (d) is magnified ($\times 2$) and offset for comparison. Peaks marked with $E_{ii}Ph$ ($i = 1, 2$) denote phonon-sideband peaks.[16-18,28] PL peaks marked with asterisks (*) in (b) and (d) are the same peak, which corresponds to an area indicated by an oval mark in (c).

Fig.3(a) and 3(b) shows the decomposed PL maps. In Fig.3 (c) and 3(d) we show the decomposed PLE spectra of (7, 5) and (6, 5) SWNTs. While the main peaks corresponding to $\Delta\mu = 0$ transitions are seen in the $I_{//}$ spectra in Fig.3(a) and 3(c), it can be clearly seen that the $I_\perp$ spectra in Fig. 3(b) and 3(d) exhibit two distinct peaks below the respective $E_{22}$ peaks, and there exist small, but nonzero intensity tails above the distinct peaks in the $I_\perp$ maps. The lowest $I_\perp$ peaks are relatively sharp, while the second $I_\perp$ peaks are broader. Although the peak positions of the second $I_\perp$ peak of (7, 5) SWNTs may be ambiguous due to the overlap of the large $E_{22}$ peaks, the peak for (7, 5) SWNTs marked with asterisk (*) in Fig.2 can be clearly seen in the PLE spectra for $I_\perp$. We confirmed that the ambiguity of $r_{//}$ around 0.3 (0.28 ~ 0.32) does not substantially change the peak positions of the lowest $I_\perp$ peaks. For (6, 5) and (7, 6) SWNTs, the peak positions of second $I_\perp$ peaks were also confirmed to be robust.

Fig.4 shows the measured PL maps, anisotropy map and decomposed PL maps for SWNTs with larger diameters than (6, 5) and (7, 5) SWNTs shown in Fig.3. As is the case for (6, 5) and (7, 5) SWNTs, the values of $r_{exp}$ at the $E_{22}$ energies were almost the same (~0.3) for each (n, m) type as shown in Fig.4 (c), and in Fig.4 (e) it is also seen that the $I_\perp$ spectra exhibit two distinct peaks below the respective $E_{22}$ peaks. This result suggests that a pair of peaks for the perpendicular photoexcitation is not a special case for a specific (n, m) type.

**D. Comparison of observed peak positions for parallel and perpendicular excitations**

Fig.5 compares the observed peak positions for $I_{//}$ and $I_\perp$ spectra with the calculated peak positions for parallel and perpendicular excitations within the geometry optimized tight-binding (TB) calculation including curvature effects.[29] It is seen that the pairs of peaks for perpendicular polarization can be expected from the TB calculation shown in Fig.5(b). When the electron-hole asymmetry is negligible, the $E_{12}$ and $E_{21}$ band gap energies are degenerate and correspond with the average value of the $E_{11}$ and $E_{22}$ energies $(E_{11} + E_{22})/2$.[24] However, when the asymmetry between valence and conduction bands is taken into account, the $E_{12}$ and $E_{21}$ energies are no longer degenerate.[24] Hence, the observed pairs of peaks in the $I_\perp$ spectra could be attributed to the pairs of $E_{12}$ and $E_{21}$ ($\Delta\mu = \pm 1$) transitions. The energy differences between the observed pairs of peaks in the $I_\perp$ spectra are approximately 100 meV for each (n, m) type, which are larger than but comparable to the TB result as shown in Fig. 5.

Comparing the peak positions for parallel and perpendicular excitations, the relative peak positions are qualitatively different between the measured results and calculation by a single particle theory. It can be clearly seen in Fig.5(a) that the energies of perpendicular excitations are considerably blue-shifted (by about 200~300 meV) from ($E_{11}$ + $E_{22}$)/2, which is the qualitative value of the transition energy for perpendicular photoexcitations predicted within a single-particle theory even when the asymmetry between valence and conduction bands is taken into account as shown in Fig.5(b). A similar blue-shift of excitonic $E_{12}$ and $E_{21}$ transitions has been reported by Zhao and Mazumdar[30] for the case of degenerate $E_{12}$ and $E_{21}$ excitations by correlated-electron calculations. In their report, they have mentioned that the amount of blue shift depends on the strength of Coulomb interaction in SWNTs. On the other hand, Uryu and Ando[31] have recently reported a large blue-shift of an excitonic absorption peak for perpendicular polarization based on theoretical calculations taking into account the depolarization effect[2]. In their calculation, the amount of the blue shift also depended on the strength of the Coulomb interaction. Assuming the amount of the blue-shift from ($E_{11}$ + $E_{22}$)/2 roughly corresponds to the difference of average exciton binding energies for ($E_{11}$, $E_{22}$) and ($E_{12}$, $E_{21}$) photoexcitations, we can attribute the observed blue-shift of the $I_\perp$ peaks to the smaller exciton binding energies for $E_{12}$ and $E_{21}$ photoexcitations. Although the theoretical calculations mentioned above[30,31] are based on different schemes, both results indicate that the amount of the blue shift depends on the strength of Coulomb interaction in SWNTs. Therefore, our result may give a measure of the strength of Coulomb interactions that have never been elucidated.

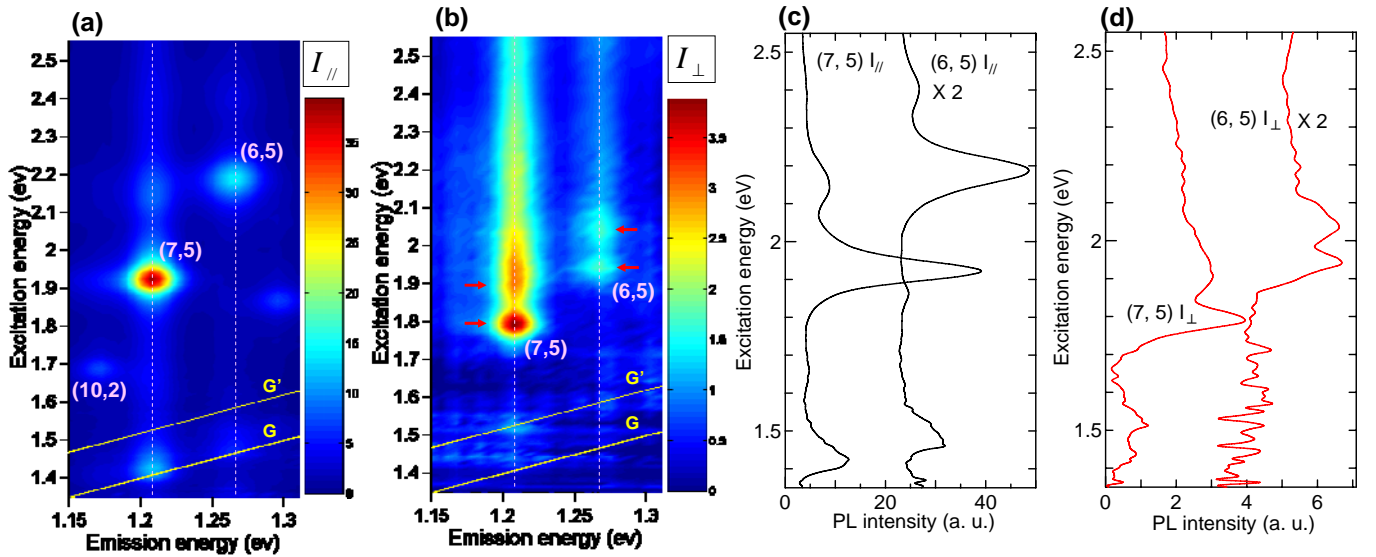

FIG. 3. (Color online) Decomposed PL maps for (a) collinear ($I_{//}$) and (b) perpendicular ($I_\perp$) dipoles. Dotted lines in (a) and (b) indicate the emission energies of respective SWNTs. Solid lines indicate the position of Raman lines for G and G' bands.[1] Peaks for $I_\perp$ spectra are indicated by arrows. Decomposed PLE spectra of (6, 5) and (7, 5) SWNTs for (c) collinear ($I_{//}$) and (d) perpendicular ($I_\perp$) dipoles. The PLE spectra for (6, 5) SWNTs are magnified ($\times 2$) and offset for comparison in (c) and (d).

Even though we have attributed the origin of a pair of peaks for perpendicular excitation to non degenerate $E_{12}$ and $E_{21}$ transitions, there is also another possibility of interpretation. According to the recent calculation by Uryu and Ando[31,32] mentioned above, when the Coulomb interaction in SWNTs is strong, the calculated spectra for the perpendicular excitation showed another distinct peak corresponding to an excited exciton state below the lowest band edge in addition to the largest peak for the first exciton state.

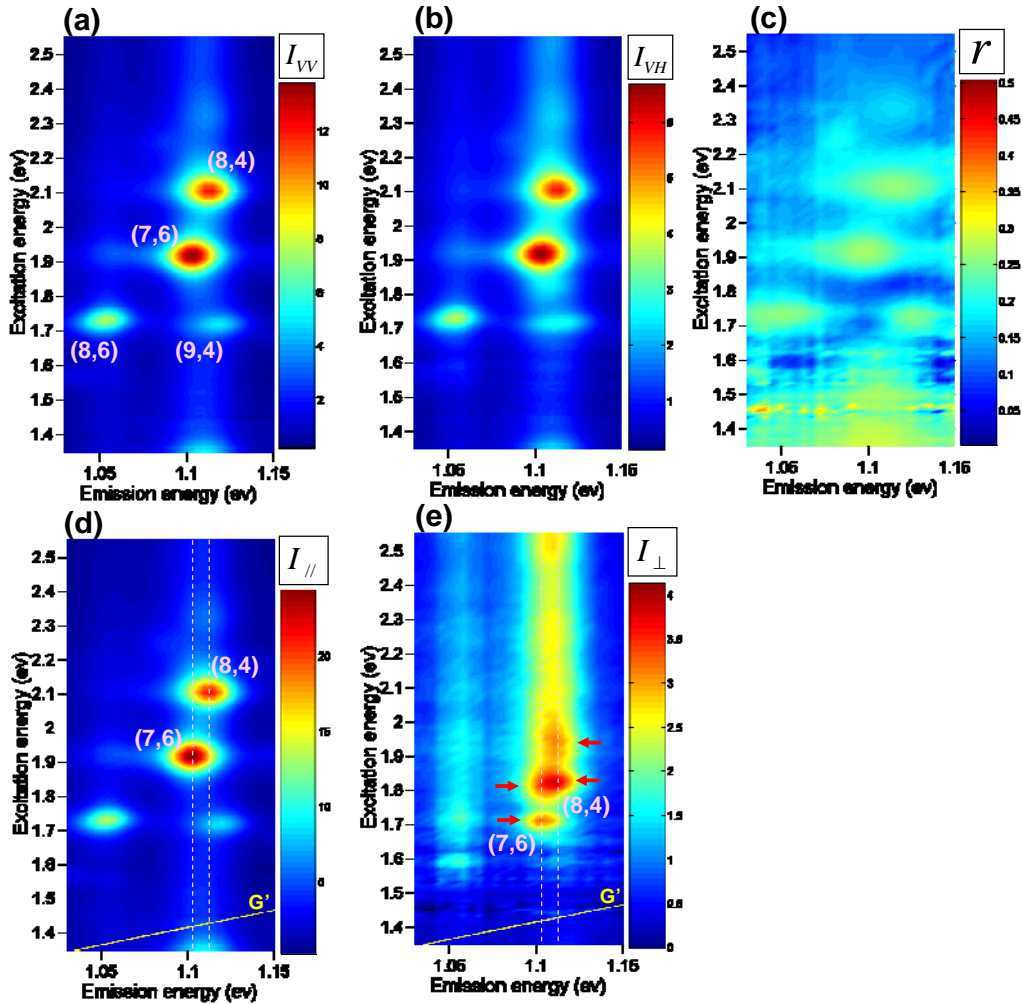

FIG. 4. (Color online) (a, b) PL and (c) anisotropy maps in the emission energy range of 1.03-1.15 eV. Decomposed PL maps for (d) collinear ($I_{//}$) and (e) perpendicular ($I_\perp$) dipoles. Dotted lines in (d) and (e) indicate the emission energies of respective SWNTs. Solid lines indicate the position of Raman lines for G' band.[1] In Fig.4(e), PL peaks for $I_\perp$ spectra are indicated by arrows.

## IV. SUMMARY

In summary, we performed anisotropic PLE measurements on SWNTs in aqueous suspension for the UV-VIS-NIR range. We found that some PL peaks for cross-polarized excitation can be clearly observed in the PLE spectra of micelle-suspended SWNTs. In addition, the observed transition energies for perpendicular excitations are blue-shifted compared with the approximate value predicted by a simple theory that does not consider excitonic effects. This qualitative discrepancy between the experimental results and the theory indicates a smaller exciton binding energy for perpendicular photoexcitations than those for $E_{11}$ and $E_{22}$ transitions.

## ACKNOWLEDGMENTS

The authors are grateful to E. Einarsson (The University of Tokyo), S. Uryu and T. Ando (Tokyo Institute of Technology), R. Saito (Tohoku University), S. Reich (Massachusetts Institute of Technology), V. Perebeinos (IBM Research Division), M.M. Kappes (Universität Karlsruhe), and G. Dukovic (Columbia University) for valuable discussions. Parts of this work were financially supported by KAKENHI No. 16360098 from JSPS. One of the authors (YM) is supported by a JSPS Research Fellowships for Young Scientists (No. 16-11409).

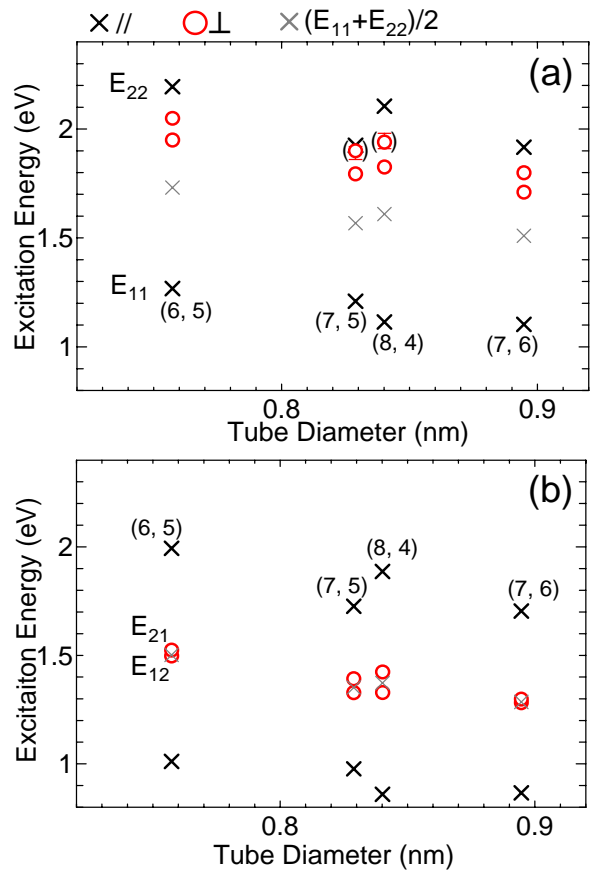

FIG. 5. (Color online) (a) Observed peak positions in PLE spectra for several (n, m) types. Symbols in parentheses are ambiguous by approximately ±40 meV due to overlap of other large PL peaks. (b) Calculated transition energies for $\Delta\mu = 0$ (black cross) and $\Delta\mu = \pm 1$ (red circle) transitions within TB approximation.